\documentclass[twocolumn]{jpsj2} %% two-column layout
\title{The entanglement of the $XY$ spin chain in a random magnetic field}

\author{Masashi \textsc{Fujinaga}$^{1}$ and Naomichi \textsc{Hatano}$^{2}$\thanks{Corresponding author. E-mail: hatano@iis.u-tokyo.ac.jp}}

\inst{$^{1}$Department of Physics, University of Tokyo, Komaba, Meguro, Tokyo 153-8505\\
$^{2}$Institute of Industrial Science, University of Tokyo, Komaba, Meguro, Tokyo 153-8505}

\abst{We investigate the entanglement of the ferromagnetic $XY$ model in a random magnetic field at zero temperature and in the uniform magnetic field at finite temperatures. We use the concurrence to quantify the entanglement. We find that,  
in the ferromagnetic region of the uniform magnetic field $h$, all the concurrences are \textit{generated} by the random magnetic field and by the thermal fluctuation. In one particular region of $h$, the next-nearest neighbor concurrence is generated by the random field but not at finite temperatures. We also find that the qualitative behavior of the maximum point of the entanglement in the random magnetic field depends on whether the variance of its distribution function is finite or not. }

\kword{entanglement, concurrence, entanglement of formation, thermal entanglement, decoherence, random field, $XY$ model}

\begin{document}
\maketitle
%%%%%%%%%%%%%%%%%%%%%%%%%%%%%%%%%%
\section{Introduction} %% No sections necessary for express letters, letters and short notes
We study in the present paper the pairwise entanglement of the ferromagnetic spin-1/2 $XY$ chain in a random magnetic field at zero temperature and in the uniform magnetic field at finite temperatures. The entanglement~\cite{sch,epr} is one of the most interesting features of quantum mechanics. It has the property of non-locality originating in the principle of superposition. One typical example of the state which has the property of the entanglement is the singlet $|\psi\rangle=1/\sqrt{2} (|01\rangle-|10\rangle)$. %Such entangled state violates the Bell inequality (CHSH inequality).
The measurement on one particle of this state affects the other particle immediately, even if two particles are far away from each other.  
This non-locality of the entangled state puzzled many people including Einstein in the early times when the quantum mechanics was born; Einstein thought that quantum mechanics was an incomplete theory because of the non-locality. However, violation of the Bell inequality~\cite{bell}, or more generally the CHSH inequality~\cite{chsh}, showed that the non-locality is a reality. Quantum mechanics  is now widely accepted, having explained a lot of phenomena which were not explained by classical mechanics.  %Thus, the entanglement is absolutely quantum; classical mechanics does not have the entanglement. 

The quantum information processing~\cite{nielsen} such as quantum teleportation~\cite{teleport} and super dense coding~\cite{dense} have been heavily studied recently. The entanglement is vital for implementation of such techniques. 
In reality, however, the entanglement may be destroyed by some decoherence effects~\cite{nielsen} such as the thermal fluctuation and a random magnetic field. Thus, it is important to know how the entanglement is affected by thermal and impurity disturbances. Effects of impurities on the entanglement are also interesting from the viewpoint of the relation between quantum coherence and impurities. 

In the present paper, we calculate the entanglement between two spins of the ferromagnetic isotropic $XY$ chain in a random magnetic field as well as at finite temperatures.
To our knowledge, this is the first to study systematically the dependence of the entanglement on the randomness in a spin system.
Li \textit{et al.}~\cite{Li} studied the dependence in the one-electron Anderson model in one dimension.
The conclusion that the entanglement \textit{increases} due to the randomness in some parameter regions (see below) is common to both studies.
(After submitting the present paper, we noticed a study on the entanglement in random quantum spin-$S$ chains.~\cite{Saguia}
It, however, is not quite related to the present issue; the study concerns a scaling law of the entanglement entropy in the random singlet phase.)

There have been a couple of works which studied the effects of the temperature on the entanglement of spin systems. Arnesen \textit{et al.}~\cite{heisent} mentioned that the nearest-neighbor entanglement of the anti-ferromagnetic Heisenberg chain can be increased by introducing the temperature in a uniform magnetic field. Similar work has been done by Nielsen~\cite{Dnielsen} on the two-spin Heisenberg model. Osborne and Nielsen~\cite{osborne} studied the nearest-neighbor and the next-nearest-neighbor entanglement of the anisotropic $XY$ chain and the ferromagnetic transverse Ising chain. Although their main interest is in the entanglement near the quantum ground-state phase transition, they also mentioned calculation of the entanglement at finite temperatures. Yano and Nishimori~\cite{yano1,yano2} also mentioned a finite-temperature calculation of the nearest-neighbor entanglement on the anti-ferromagnetic anisotropic $XY$ model. Their results for the nearest-neighbor entanglement are almost the same as ours.
Our conclusion that the entanglement also increases due to the thermal fluctuation in some parameter regions (see below) is common to the above-mentioned studies.
None, however, compared the entanglement in a random magnetic field and that at finite temperatures quantitatively in the same model.

We compute the entanglement between two spins from the nearest-neighbor pair to the fifth-neighbor pair. We use the concurrence~\cite{wootters} to quantify the pairwise entanglement. 
%In the ferromagnetic $XY$ model at finite temperatures, the nearest-neighbor concurrences are almost the same as that of the anti-ferromagnetic $XY$ model~\cite{yano1} at finite temperatures, whereas the concurrences of the ferromagnetic Heisenberg model has no entanglement at finite temperatures~\cite{heisent}. 
%In a random magnetic field, we will defined the average concurrence in \S~\ref{part3} and use it.
We find the following: 
\begin{itemize}
\item[(1)] In general, the entanglement is decreased as the randomness is increased; 
\item[(2)] In the region of the uniform magnetic field $h>1$, the entanglement is increased by the random magnetic field and by the temperature.
In yet another region $h<1/2$, it is increased by the random field but not by the temperature.
\item[(3)] Qualitative behavior of the maximum point of the entanglement depends on the random magnetic field, in particular, whether the variance of the distribution function is finite or not.
\item[(4)] The entanglement between two spins at finite temperatures becomes weaker than that in the random magnetic field at zero temperature as the distance between the two spins gets greater.
\end{itemize}
The present paper is organized as follows. In \S\ref{ppart}, we introduce the model and review the outline of computation of its correlation functions, which are necessary to quantify the concurrence. In \S\ref{numericalresults}, we calculate the concurrence for: in \S\ref{uni-zero-temp}, the $XY$ spin chain in the uniform magnetic field at zero temperature; in \S\ref{ran-zero-temp}, the $XY$ spin chain in a random magnetic field at zero temperature; in \S\ref{uni-finite-temp}, the $XY$ spin chain in the uniform magnetic field at finite temperatures. Finally, we give a summary and discussions in \S\ref{part4}.
%%%%%%%%%%%%%%%%%%%%%%%%%%%%%%%%%%%%
\section{The model and the entanglement} \label{ppart}
\subsection{Diagonalization of the model Hamiltonian} \label{part2}
The Hamiltonian of the $XY$ spin chain in a random magnetic field is given in the form
\begin{equation}
	H=-\frac{J}{4} \sum_{j=1}^N \left(\sigma^x_j \sigma^x_{j+1} + \sigma^y_j \sigma^y_{j+1}\right) - \frac{1}{2}\sum_{j=1}^N \left(h+h_j\right) \sigma^z_j, \label{model}
\end{equation}
where $J \ (>0)$ is the coupling constant, $N$ is the number of the spins, $\sigma^\alpha$ $(\alpha= x,y,z)$ are the Pauli matrices,
%\begin{equation}
%	\sigma^x = \begin{pmatrix}
%			 0 & 1 \\
%			 1 & 0 \notag
%			 \end{pmatrix}, \quad
%	\sigma^y = \begin{pmatrix}
%			 0 & -\textrm{i} \\
%			 \textrm{i} & 0 \notag
%			\end{pmatrix},  \quad	
 %	\sigma^z = \begin{pmatrix}
%			1 & 0 \\
%			0 & -1 \notag
%			\end{pmatrix}, 		
%\end{equation}
$h$ is the uniform magnetic field and $\{h_j\}$ are the random magnetic field. We impose the periodic boundary conditions: 
\begin{equation}
	\sigma^\alpha_{N+1}=\sigma^\alpha_1, \quad (\alpha=x,y,z).  \label{periodic}
\end{equation}
Hereafter, the coupling constant $J$ is set to one. 
The random magnetic field $h_j$ at each site obeys the distribution function
\begin{equation}
	P_{q,a}\left(h_j\right) \sim \left[a^2-(1-q){h_j}^2 \right]^{\frac{1}{1-q}}, \label{probfunction}
\end{equation}
where the parameter $q$ determines the type of the distribution function and $a$ determines the width of the distribution function. In particular, eq.~\eqref{probfunction} is reduced to a Gaussian distribution function as $q \rightarrow 1$. In this case, the scale parameter $a$ is its standard deviation. Equation~\eqref{probfunction} is also reduced to a Lorentzian distribution function for $q=2$. In this case, the scale parameter $a$ is its half width at half maximum. 
The variance of the distribution function~\eqref{probfunction} diverges for $q\geq 5/3$ and is finite for $q<5/3$.
In the case of the Lorentzian distribution $q=2$, Nishimori~\cite{nishimori1} analytically calculated the average one-point correlation function and obtained lower bounds of the average two-point correlation functions. The results in the paper~\cite{nishimori1}, however, are not used in the present paper, since we take the random average of the concurrence, which is a non-linear function of the one-point and two-point correlation functions.

We diagonalize the Hamiltonian~\eqref{model} as follows. The Hamiltonian can be expressed by the Fermi operators $a^\dagger$ and $a$ after the Jordan-Wigner transformation.
%\begin{align}
%	\sigma^+_j &= \prod^{j-1}_{l=1} \exp(-\textrm{i} \pi a^\dagger_l a_l) a^\dagger_j, 
%	\\	
%	\sigma^-_j &= \prod^{j-1}_{l=1} \exp(\textrm{i} \pi a^\dagger_l a_l) a_j,  \\
%	\sigma^z_j &= 2\sigma^+_j \sigma^-_j = 2a^\dagger_j a_j -1,  
%\end{align}
%where $a^\dagger_l$ and $a_l$ satisfy the anti-commutation relations $\{a^\dagger_l, a_m\}=\delta_{lm}$ and $\{a_l, a_m\}=0$.
The Hamiltonian is then reduced to the quadratic form
\begin{equation}
	H=\sum_{i,j=1}^{N} a^\dagger_i A_{ij} a_j 
	\label{hamiltonian}
\end{equation}
with 
%\begin{equation}
%	A = 
%	\begin{pmatrix}
%-h-h_1      & -\frac{1}{2}& 0           & \cdots     & 0          &\pm \frac{1}{2} \\
%-\frac{1}{2}& -h-h_2      & -\frac{1}{2}& 0          & \vdots     &    0 \\
%0           & -\frac{J}{2}& \ddots      & \ddots     &0           &\vdots \\
%\vdots      &   0         & \ddots      & \ddots     &-\frac{1}{2}& 0 \\  
%0           & \vdots      &  \ddots     &-\frac{1}{2}&-h-h_{N-1}  &-\frac{1}{2}  \\
%\pm \frac{1}{2}& 0        &\cdots       &      0     &-\frac{1}{2}& -h-h_N
%	\end{pmatrix}, \label{matrixform}
%\end{equation}
\begin{eqnarray}
&&
A =
\nonumber\\
&&\left(\begin{array}{ccccc}
-h-h_1      & -\frac{1}{2}& 0           & \cdots        &\pm \frac{1}{2} \\
-\frac{1}{2}& -h-h_2      & -\frac{1}{2}& \ddots              &    0 \\
0           & -\frac{1}{2}& \ddots      & \ddots         &\vdots \\
\vdots      &   \ddots         & \cdots      & \ddots & -\frac{1}{2}  \\
\pm \frac{1}{2}& 0       &\cdots       & -\frac{1}{2} & -h-h_N
	\end{array}\right),
\nonumber\\
	\label{matrixform}
\end{eqnarray}
where we dropped a constant term in the Hamiltonian~\eqref{hamiltonian}. The signs of the $(1,N)$ and $(N,1)$ elements in eq.~\eqref{matrixform} are negative when the number of the Fermions in the system is even and positive when odd. 
The Hermitian matrix $A$ is diagonalized by a unitary matrix $V$. 
%\begin{equation}
%V^\dagger A V=\Lambda \label{Vmatrix}
%\end{equation}
%is a diagonal matrix. 
We thus have 
%the Hamiltonian~\eqref{hamiltonian} without the second term of the right hand side in eq.~\eqref{hamiltonian} as
\begin{equation}
	H=\sum_{i=1}^N \epsilon_ic^\dagger_{i} c_i, \label{dhamiltonian}
\end{equation}
where the operators $c^\dagger_i$ and $c_i$ are given by
\begin{equation}
	c^\dagger_i =\sum_{l=1}^N a^\dagger_l V_{li}, \quad
	c_i = \sum_{l=1}^N V^\dagger_{il}a_l
\end{equation}
and satisfy the anti-commutation relations $\{c^\dagger_i, c_j\}=\delta_{ij}$ and $\{c_i, c_j\}=0$.

\subsection{Correlation functions and the two-site density matrix}
In order to quantify the entanglement, we use the concurrence~\cite{wootters} related to the entanglement of formation~\cite{bennet}. The concurrence between the spins at sites $i$ and $j$ is calculated from the two-site density matrix $\rho_{ij}$ as
\begin{equation}
	C_{i,j}= \textrm{max}\left\{0, \lambda_1-\lambda_2-\lambda_3-\lambda_4\right\},
\end{equation}
where $\{\lambda_i\}_{i=1}^4$ are the square roots of the eigenvalues of the matrix $R=\rho_{ij} {\tilde{\rho}}_{ij}$ in non-ascending order, $\lambda_1 \geq \lambda_2 \geq \lambda_3 \geq \lambda_4$ with
${\tilde{\rho}}_{ij}=\left(\sigma^y_i \otimes \sigma^y_j \right)\rho^* \left(\sigma^y_i \otimes \sigma^y_j \right)$. The complex conjugation is taken in the 
$\sigma^z$ basis.
The two-site density matrix $\rho_{ij}$ is defined by
\begin{equation}
	\rho_{ij}=\textrm{Tr}_{\Hat{ij}}\rho,
\end{equation}
where $\textrm{Tr}_{\Hat{ij}}$ denotes the trace over the degrees of freedom except for the sites $i$ and $j$, and $\rho$ is the density matrix of the whole system: $\rho=e^{-\beta H}/Z$. 

The two-site density matrix can be expanded in terms of the identity matrix and the Pauli matrices as
\begin{equation}
	\rho_{ij}= \frac{1}{4}\sum_{\alpha,\beta=0}^3 p_{\alpha \beta}
	\sigma^\alpha_i \otimes \sigma^\beta_j,  \label{proddensity}
\end{equation}
where $\sigma^0_i$ denotes the identity operator on the site $i$, $\sigma^1_i=\sigma^x_i$, $\sigma^2_i=\sigma^y_i$ and $\sigma^3_i=\sigma^z_i$.
The coefficients $p_{\alpha \beta}$ are real numbers determined by
\begin{equation}
	p_{\alpha \beta}=\textrm{Tr}\left(\sigma^\alpha_i \sigma^\beta_j \rho_{ij}\right)= \langle \sigma^\alpha_i \sigma^\beta_j \rangle.
\end{equation}
Hence, sixteen coefficients are needed to determine the two-site density matrix in general. Thanks to the symmetry of the Hamiltonian~\eqref{model}, the number of the coefficient is reduced to four; we need $\langle\sigma^z_i\rangle$, $\langle\sigma^z_i\rangle$ and $\langle\sigma^\alpha_i\sigma^\alpha_j\rangle$ $(\alpha=1,3)$ only. The others are zero~\cite{osborne}.
%
%The density matrix of the $XY$ spin chain at thermal equilibrium is given by the canonical ensemble
%\begin{equation}
%	\rho=\frac{e^{-\beta H}}{Z},
%\end{equation}
%where $\beta^{-1}=k T$, $k$ is the Boltzmann constant and $Z$ is the partition function $\textrm{Tr}e^{-\beta H}$.
%First, the Hamiltonian~\eqref{model} is invariant under the $\pi/2$ rotation along the $z$-axis;
%$\left[H,U_{\pi/2}\right]$,
%where
%\begin{equation}
%	U_{\pi/2} = \prod_{i=1}^N \exp\left(-\textrm{i}\frac{\sigma^z_i}{4}\right).
%\end{equation}
%This means $[\rho, U_{\pi/2}]$. We thus have $\langle \sigma^x_i \sigma^x_j \rangle
%=\langle \sigma^x_i \sigma^x_j \rangle$. Second, the Hamiltonian possesses the global phase flip symmetry 
%\begin{equation}
%	U_{\textrm{pf}}=\prod_{i=1}^N \sigma^z_i.
%\end{equation}
%This means that two-point correlation functions including an odd number of $\sigma^x$ or $\sigma^y$ vanish similarly to the above arguments.
%The operator $U_\textrm{pf}$ commutes the Hamiltonians $[H,U_\textrm{pf}]$ and thus commutes the density matrix $\rho$: $[H,\rho]$. This means that two-point correlation functions including an odd number of $\sigma^x$ or $\sigma^y$ vanish.
%Finally, the Hamiltonian is a real matrix in the $\sigma^z$ basis so that 
%$\rho^*_{ij}=\rho_{ij}$. Noting that only the element in $\sigma^y$ are imaginary, we have $\langle \sigma^x_i \sigma^y_j\rangle=\langle \sigma^y_i \sigma^x_j\rangle=0$. 
Hence, the two-site density matrices of the model take the form
\begin{align}
	\rho_{ij}=\frac{1}{4} \Big(I_{ij} &+\langle\sigma^z_i\rangle\sigma^z_i\otimes I_j + \langle \sigma^z_j\rangle I_i \otimes \sigma^z_j 
	\notag \\
	&+ \sum_{\alpha=1}^3
\langle\sigma^{\alpha}_i\sigma^{\alpha}_j\rangle\sigma^{\alpha}_i\otimes\sigma^{\alpha}_j\Big),
\end{align}
where $\sigma^1=\sigma^x$, $\sigma^2=\sigma^y$, $\sigma^3=\sigma^z$ and
$\langle \sigma^1_i \sigma^1_j\rangle=\langle \sigma^2_i \sigma^2_j\rangle$.
%Therefore, $\langle\sigma^z_i\rangle$, $\langle\sigma^z_i\rangle$ and $\langle\sigma^\alpha_i\sigma^\alpha_j\rangle$ $(\alpha=1,2)$ are required to determine two-site density matrix of the model; the number of the independent coefficients are thus reduced to four. In the zero temperature, taking the limit $\beta \rightarrow \infty$ leads the same results.

The correlation functions are obtained as follows~\cite{lieb,chak,nishimori}:
\begin{align}
	\langle \sigma^z_i \rangle
	&= G_{i,i} , \label{sigmaz}
\\
	\langle \sigma^z_i \sigma^z_j \rangle
	&=
	\begin{vmatrix}
		G_{i,i} & G_{i,j}  \\
		G_{j,i} & G_{j,j}
	\end{vmatrix} , \label{sigmazz}
\\
	\langle \sigma^x_i \sigma^x_j \rangle
	&=
	\begin{vmatrix}
			 G_{i,i+1}   & G_{i,i+2}    & \cdots & G_{i,j}    \\
			 G_{i+1,i+1} & G_{i+1,i+2}  & \cdots & G_{i+1,j}  \\
			 \vdots      & \vdots       & \ddots & \vdots     \\
			 G_{j-1,i+1} & G_{j-1,i+2}  & \cdots & G_{j-1,j}
	\end{vmatrix} , \label{sigmaxx}
\end{align}
where $G_{i,j}$ are given as follows in the three cases: 
%$XY$ spin chain in the uniform magnetic field at zero temperature, $XY$ spin chain in the uniform magnetic field at finite temperature and $XY$ spin chain in a random magnetic field at zero temperature.
\begin{itemize}
\item[i)]The $XY$ spin chain in the uniform magnetic field at zero temperature.
%($\{h_i=0\}$ for all $i$ in eq.~\eqref{model}, $T=0$).
In the thermodynamic limit, $\{G_{i,j}\}$ are given by~\cite{lieb,chak,nishimori}
\begin{align}
	G_{i,i} &= 
	\left\{
	\begin{array}{ll}
	1   & \textrm{for} \ h>J ,\\
	-1 + \frac{2}{\pi} \arccos ( -\frac{h}{J})   & \textrm{for} \ h<J,
	\end{array} \label{uni1}
	\right.  
\\
	G_{i,j} &= 
	\left\{
	\begin{array}{l}
	0  \\
	\hspace*{0.2\textwidth}\textrm{for} \ h>J, \\
	\frac{2}{\pi} \frac{1}{l-m} \sin \left[ (i-j) \arccos (-\frac{h}{J}) \right]  \\
	\hspace*{0.2\textwidth}\textrm{for} \ h<J.
	\end{array}  \label{uni2}
	\right.  
\end{align}
\item[ii)]The $XY$ spin chain in the uniform magnetic field at finite temperatures.
%($\{h_i=0\}$ for all $i$ in eq.~\eqref{model}). 
In the thermodynamic limit, $\{G_{i,j}\}$ are given by~\cite{lieb,chak,nishimori}
\begin{equation}
	G_{i,j} = -\delta_{ij} + \frac{2}{\pi} \int^\pi_0 \textrm{d}\phi\frac{\cos(i-j)\phi}{1+\textrm{exp}(-\beta(J \cos \phi +h) )}. \label{thermal}
\end{equation}
\item[iii)]The $XY$ spin chain in a random magnetic field at zero temperature. In this case, $\{G_{i,j}\}$ are given by
\begin{equation}
	G_{i,j}=2\sum_{l=1}^{N_G}V_{il}V_{jl}-\delta_{ij},
\end{equation}
where $V$ is the matrix diagonalizing the matrix $A$ in eq.~\eqref{matrixform} and $N_G$ is the number of the Fermions. In the ground state, the Fermions are filled in the levels with $\epsilon_i <0$ in eq.~\eqref{dhamiltonian}.
\end{itemize}
%%%%%%%%%%%%%%%%%%%%%%%%%%%%%%%%%%%%%%%%%%%%%
\subsection{The concurrence} \label{part3}
Now that the coefficients in eq.~\eqref{proddensity} have been obtained, we can evaluate the concurrence.
We here define the average concurrence as the random and spatial average:
\begin{equation}
	C(r)=\frac{1}{N}\sum_{i=1}^N \left[C_{i,i+r}\right]_{\textrm{av}}, \label{avcon}
\end{equation}
where $[\cdots]_{\textrm{av}}$ denotes the random average, $C_{i,j}$ denotes the concurrence between the sites $i$ and $j$, and $N$ is the number of the sites.
In the absence of the random magnetic field, the Hamiltonian possesses the translational invariance and the averaging is not necessary. 
We note that the entanglement of formation~\cite{bennet} after the random average and the spatial average is always greater than that obtained by substitution of the average concurrence into the relation between the entanglement of formation $E$ and the concurrence $C$
\begin{align}
	E(C)=&-\frac{1+\sqrt{1-C^2}}{2}\log_2\left(\frac{1+\sqrt{1-C^2}}{2}\right) \notag \\
	&-\frac{1-\sqrt{1-C^2}}{2}\log_2\left(\frac{1-\sqrt{1-C^2}}{2}\right),
\end{align}
since $E(C)$ is a concave function of the concurrence $C$. 
Hereafter, we simply refer to the average concurrence as the concurrence. 

\begin{figure}
\begin{center}
\includegraphics[width=0.45\textwidth,clip]{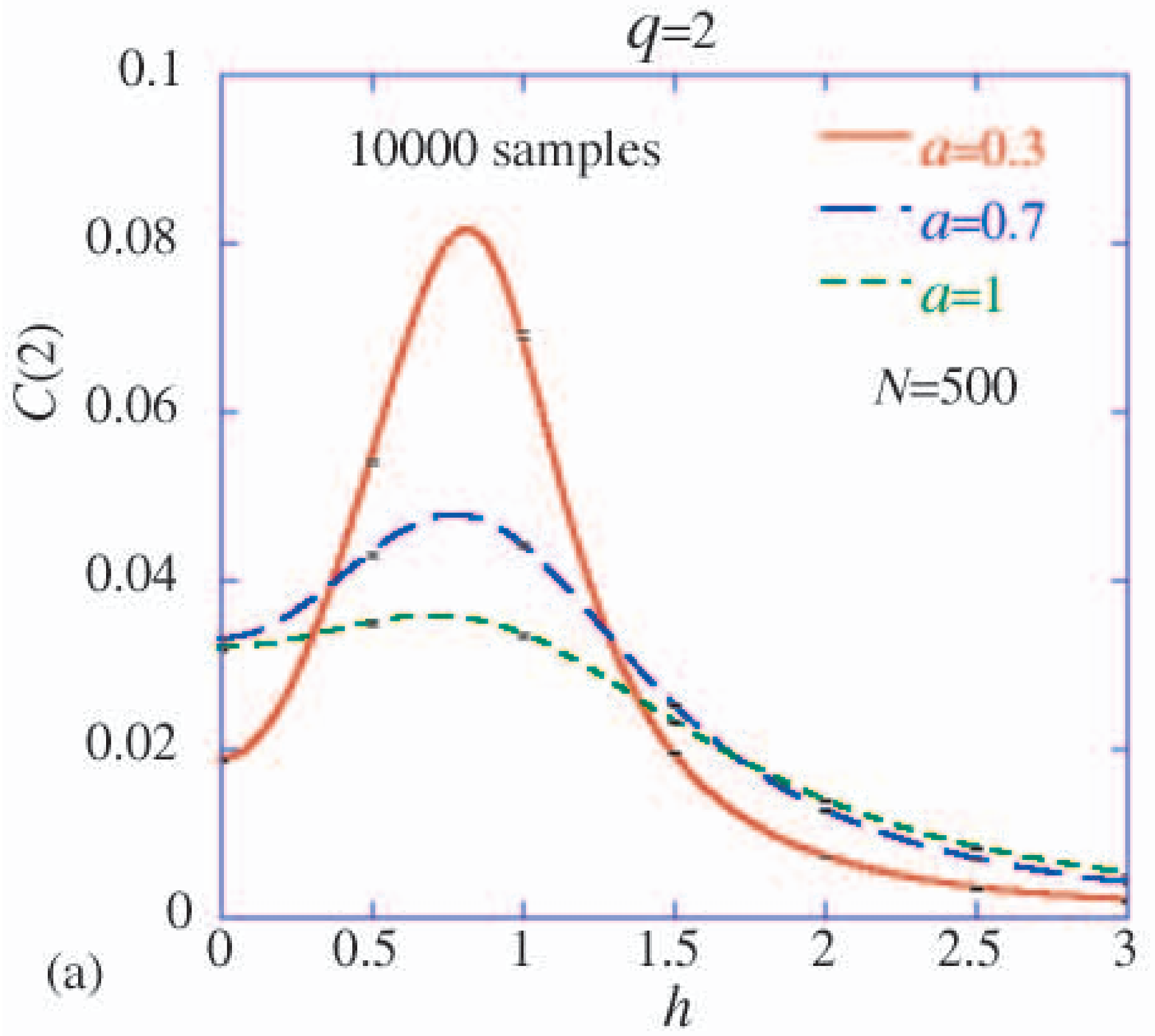}
\includegraphics[width=0.45\textwidth,clip]{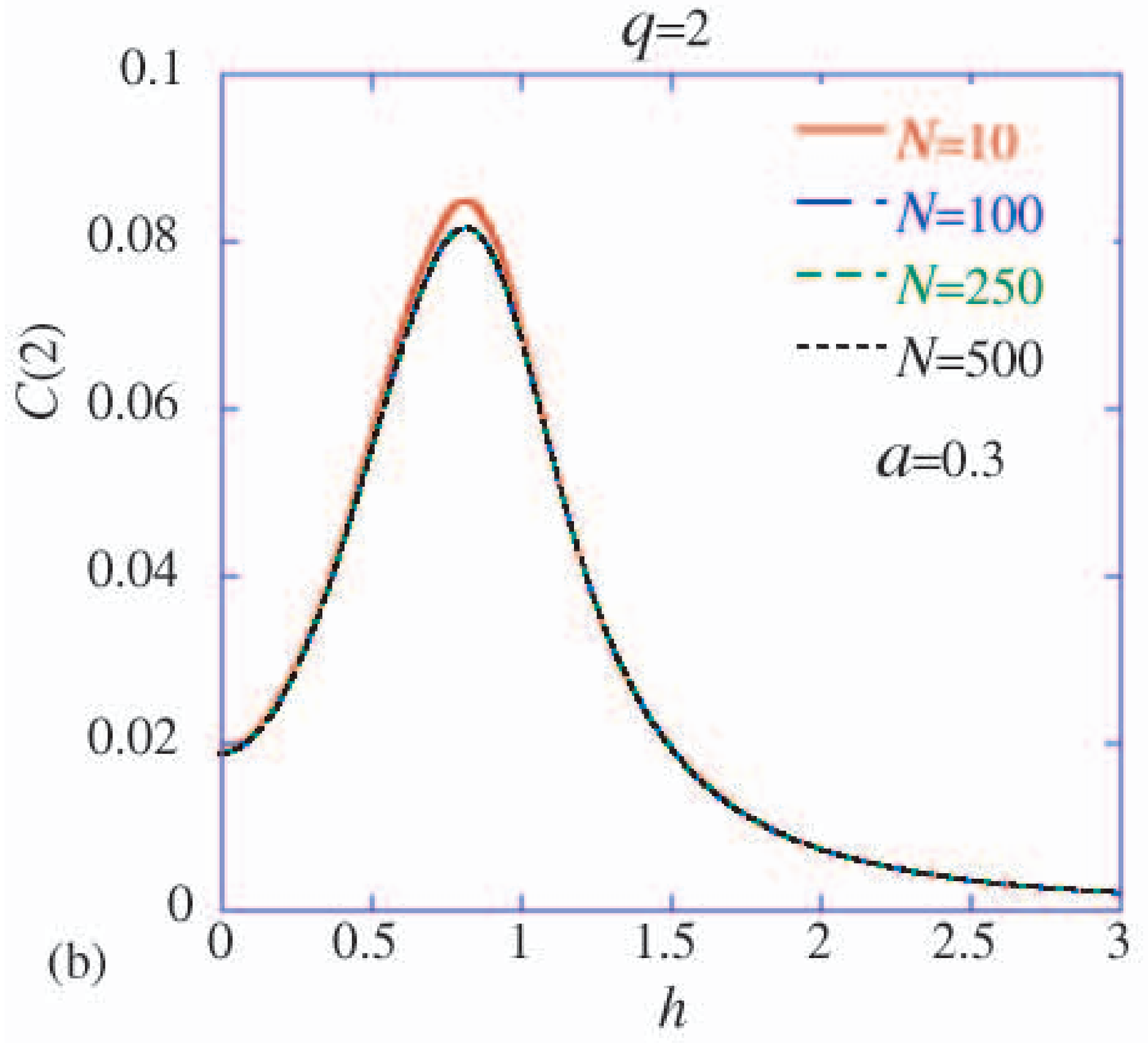}
\caption{(Color online) (a) The next-nearest-neighbor concurrence $C(2)$ for $q=2$ (the Lorentzian distribution) with the scale parameter $a=0.3$, 0.7, 1. All the lines are plotted as a function of the uniform magnetic field $h$. The random average is taken over 10000 samples with the system size 500. (b) The next-nearest-neighbor concurrence $C(2)$ for $q=2$. The lines are plotted with the scale parameter $a=0.3$ with 10000 samples and for the system size $N=10$, 100, 250, 500.}
\label{samplesize}
\end{center}
\end{figure}

In the case of no random magnetic field, we calculated the concurrence rigorously  in the thermodynamic limit.
In the case with a random magnetic field, we numerically evaluated the sample average of the concurrence~\eqref{avcon}.
For all the results below in the random case, the number of the sites $N$ is 500 and the number of the samples is 10000. In Fig.~\ref{samplesize}(a), the next-nearest-neighbor concurrence $C(2)$ is plotted for $q=2$ (the Lorentzian distribution) with error bars at $h=0$, 0.5, 1, 1.5, 2, 2.5 and 3, but the errors are almost invisible.
%%%%%%%%%%%%%%%%%%%%%%%%%%%%%%%%%%%%%%%%%%%%%%%
%%%%%%%%%%%%%%%%%%%%%%%%%%%%%%%%%%%%%%%%%%%%%%%%
In Fig.~\ref{samplesize}(b), the next-nearest-neighbor concurrence $C(2)$ is plotted for $q=2$ and for the system size $N=10$, 100, 250 and 500 with the random average over 10000 samples. The finite-size effect is invisible for $N\geq100$. We hence conclude that 10000 samples and the system size $N=500$ are substantial.
We calculated the average concurrence $C(r)$ for $1\leq r \leq 5$ but not all the results are plotted below.

%%%%%%%%%%%%%%%%%%%%%%%%%%%%%%%%%%%%%%
\section{Numerical results} \label{numericalresults}
\subsection{$XY$ spin chain in a uniform magnetic field at zero temperature}
\label{uni-zero-temp}
\begin{figure}
\begin{center}
\includegraphics[width=0.45\textwidth,clip]{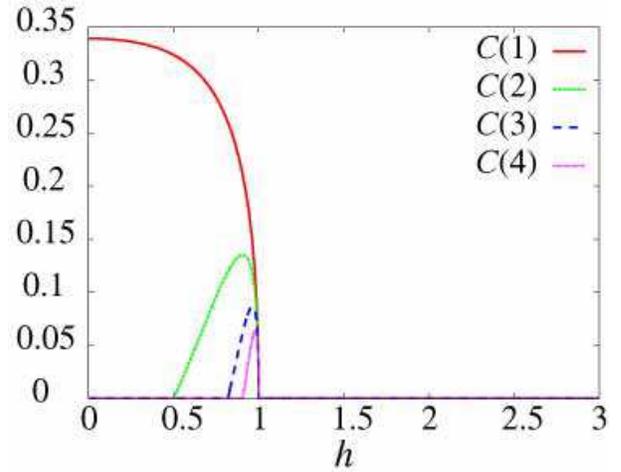}
\caption{(Color online) The concurrence of the $XY$ spin chain in a uniform magnetic field as a function of the uniform magnetic field $h$.}
\label{uni-con}
\end{center}
\end{figure}

We first study the concurrence of the $XY$ spin chain in the uniform magnetic field. 
Figure~\ref{uni-con} shows the nearest-neighbor concurrence $C(1)$, the next-nearest-neighbor concurrence $C(2)$, the third-neighbor concurrence $C(3)$ and the fourth-neighbor concurrence $C(4)$. All the concurrences rapidly decrease near
$h=1$, where the quantum phase transition occurs, and vanish in the region $h>1$. In the region $h>1$, the ground state is given by the tensor product of the one-spin state $|\uparrow_1\rangle$ as
\begin{equation}
	|\textrm{GS}\rangle =  |\uparrow_1\rangle|\uparrow_2\rangle|\uparrow_3\rangle
	\cdots , \label{gs}
\end{equation}
where the state $|\uparrow_i\rangle$ denotes the eigenstate of the matrix $\sigma^z_i$ satisfying the eigenequation $\sigma^z_i|\uparrow_i\rangle = |\uparrow_i\rangle$. Since there is no superposition involved, the entanglement vanishes in $h>1$.
%%%%%%%%%%%%%%%%%%%%%%%%%%%%%%%%%%%%%%%%%%%%%%
%%%%%%%%%%%%%%%%%%%%%%%%%%%%%%%%%%%%%%
%%%%%%%%%%%%%%%%%%%%%%%%%%%%%%%%%%%%%%%%%%%%%%
\begin{figure}
\begin{center}
\includegraphics[width=0.45\textwidth,clip]{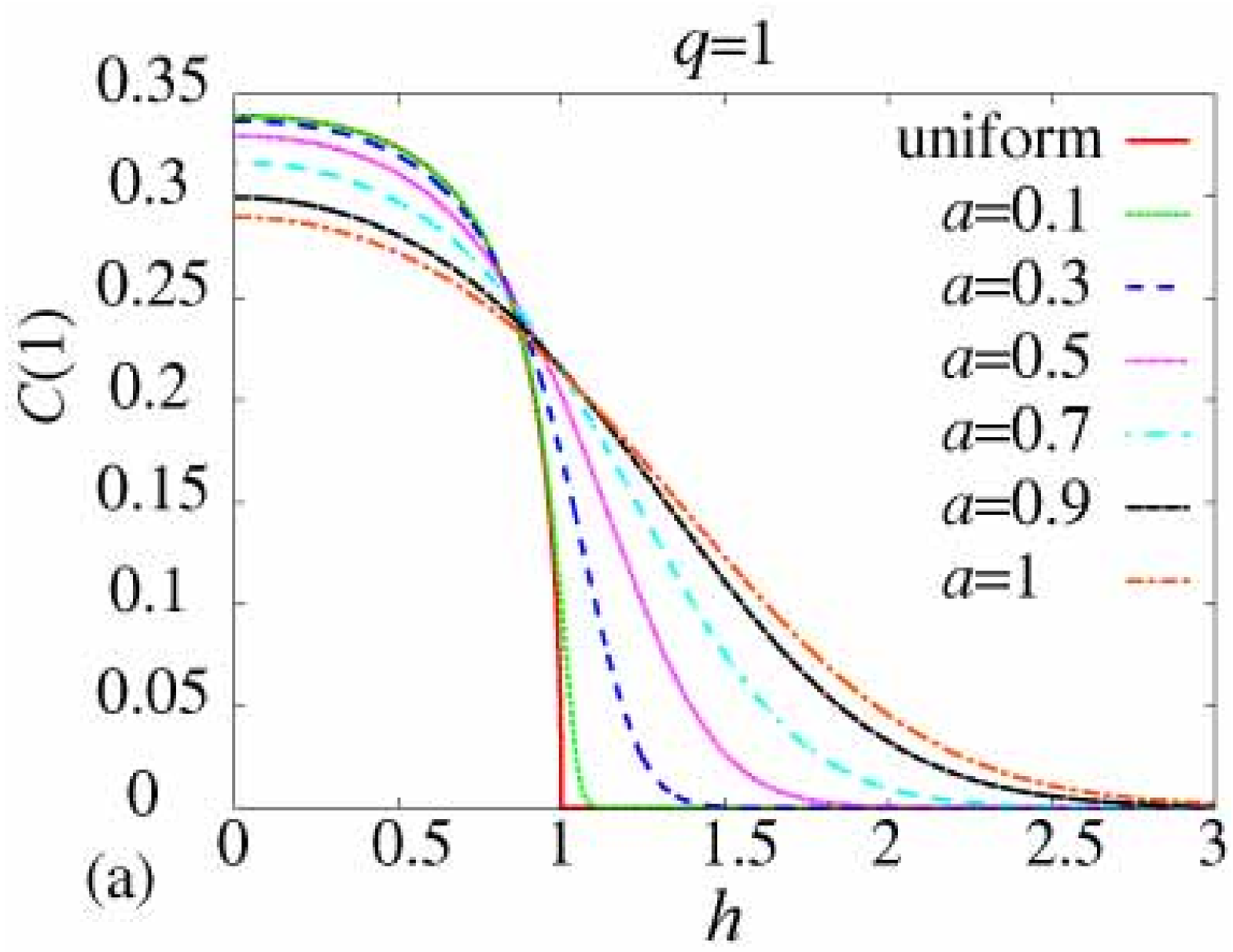} 
\includegraphics[width=0.45\textwidth,clip]{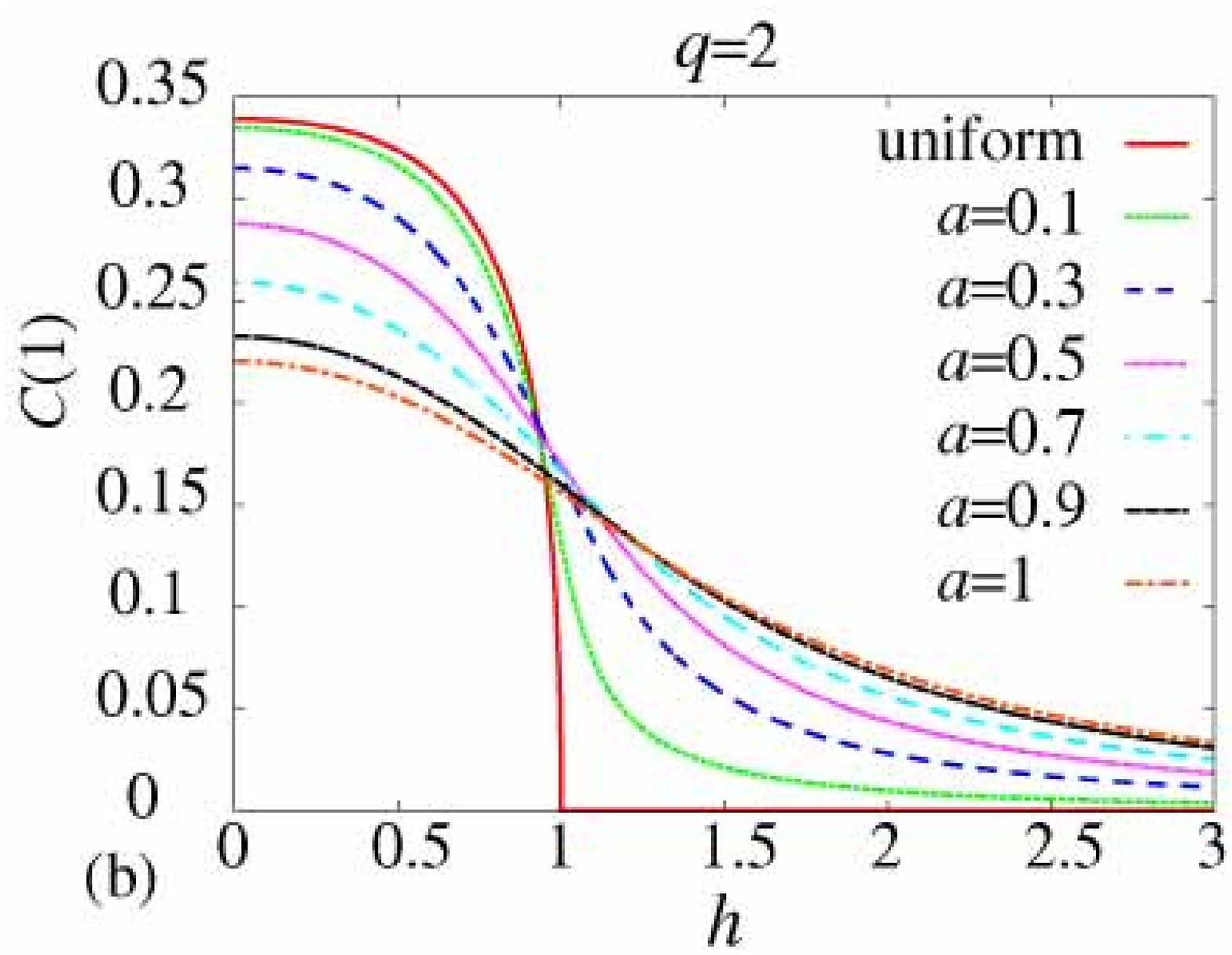}
\caption{(Color online) The nearest-neighbor concurrence $C(1)$ of the $XY$ spin chain in a random magnetic field at zero temperature; (a) for $q=1$ (the Gaussian distribution); 
%(b) for $q=1.35$;
%(c) for $q=5/3$;
%(d) for $q=1.85$;
(b) for $q=2$ (the Lorentzian distribution).
All the data are plotted as functions of the uniform magnetic field.
}
\label{c1}
\end{center}
\end{figure}
%%%%%%%%%%%%%%%%%%%%%%%%%%%%%%%%%%%%%%%%%%%%%%%%

\subsection{$XY$ spin chain in a random magnetic field at zero temperature}
\label{ran-zero-temp}
Next, we study the concurrence in a random magnetic field (in addition to the uniform magnetic field $h$) at zero temperature.
The random magnetic field obeys the distribution function~\eqref{probfunction}; we investigate the cases for $q=1$, 1.35, 5/3, 1.85 and 2.

The nearest-neighbor concurrence $C(1)$ in all cases behaves similarly. In the region $h<1$, the nearest-neighbor concurrence in a random magnetic field for each $q$ decreases as the distribution width $a$ is increased. We here show in Fig.~\ref{c1} only the cases $q=1$ and $q=2$.
The reduction of the nearest-neighbor concurrence $C(1)$ is greater as the scale parameter $a$ is increased.
For $h>1$, the nearest-neighbor concurrence for all $q$ is \textit{increased} by the random magnetic field. That is, the random magnetic field increases the quantum correlation.
The reason why the nearest-neighbor concurrence is increased for $h>1$ may be as follows; the random magnetic field flips some of the aligned spins of the ferromagnetic (or classical) ground state~\eqref{gs} and thereby the flipped spins and their neighboring spins restore the quantum interaction.
%%%%%%%%%%%%%%%%%%%%%%%%%%%%%%%%%%%%%%%%
\begin{figure*}[t]
\begin{center}
\includegraphics[width=0.45\textwidth,clip]{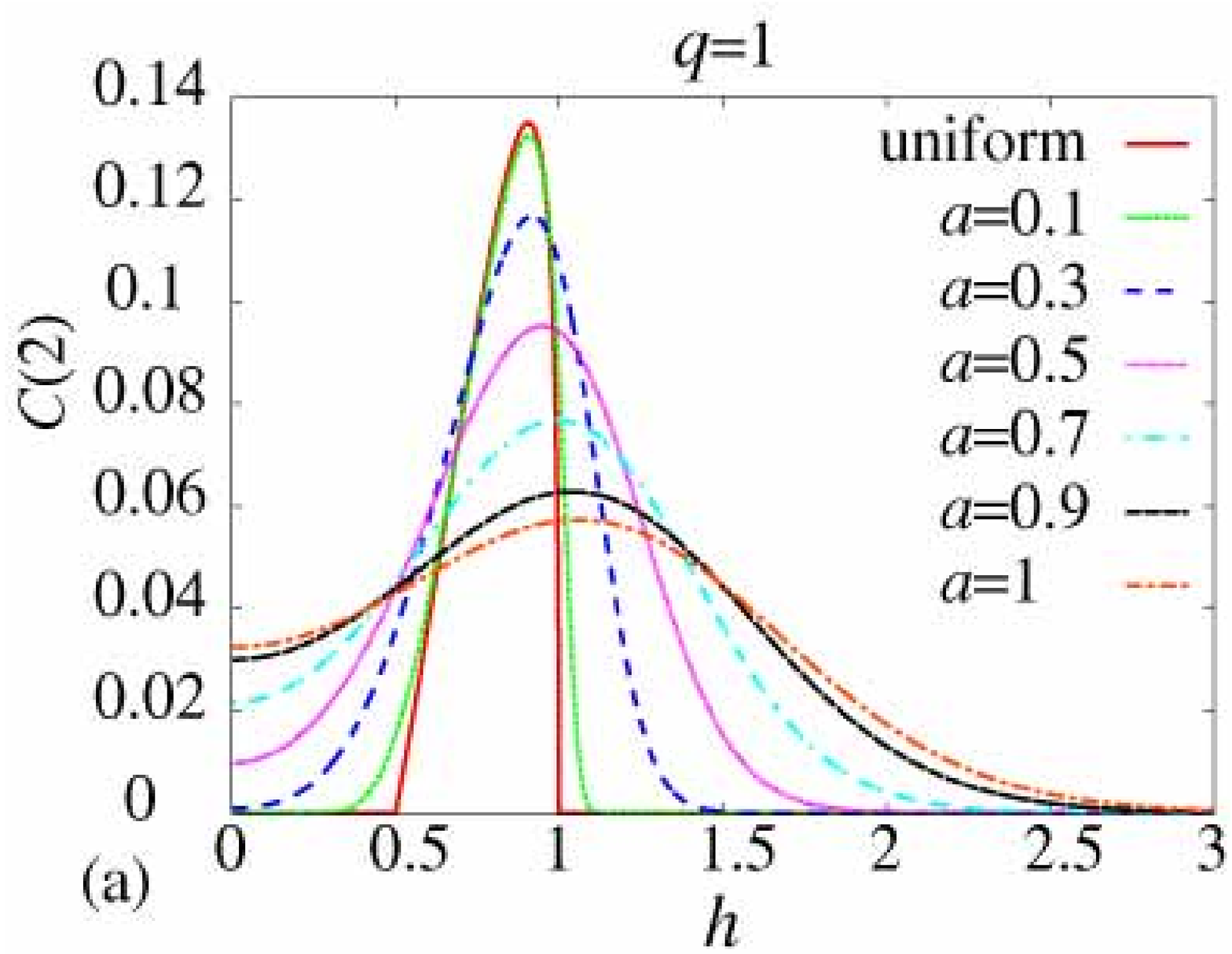} 
\includegraphics[width=0.45\textwidth,clip]{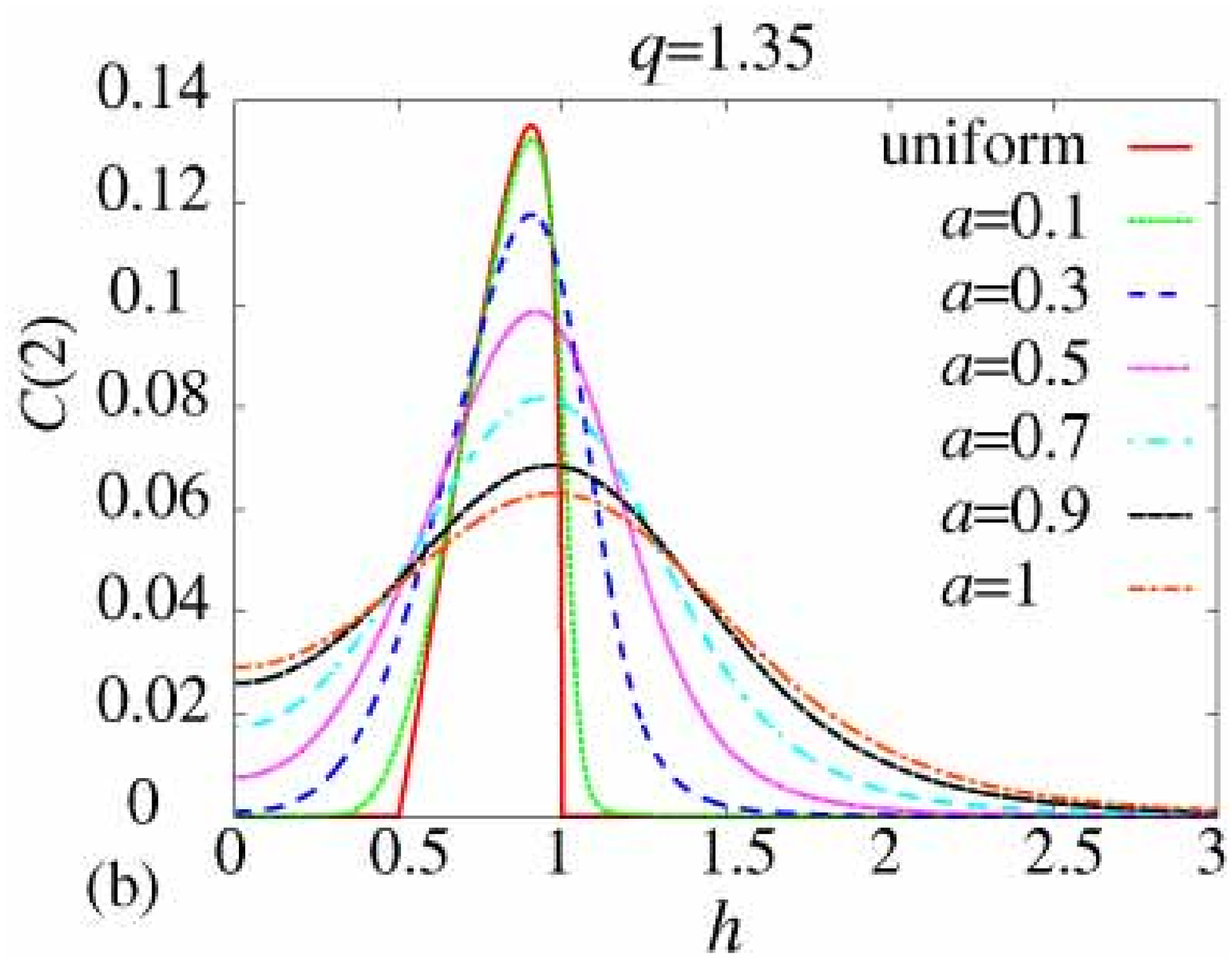}
\\
\includegraphics[width=0.45\textwidth,clip]{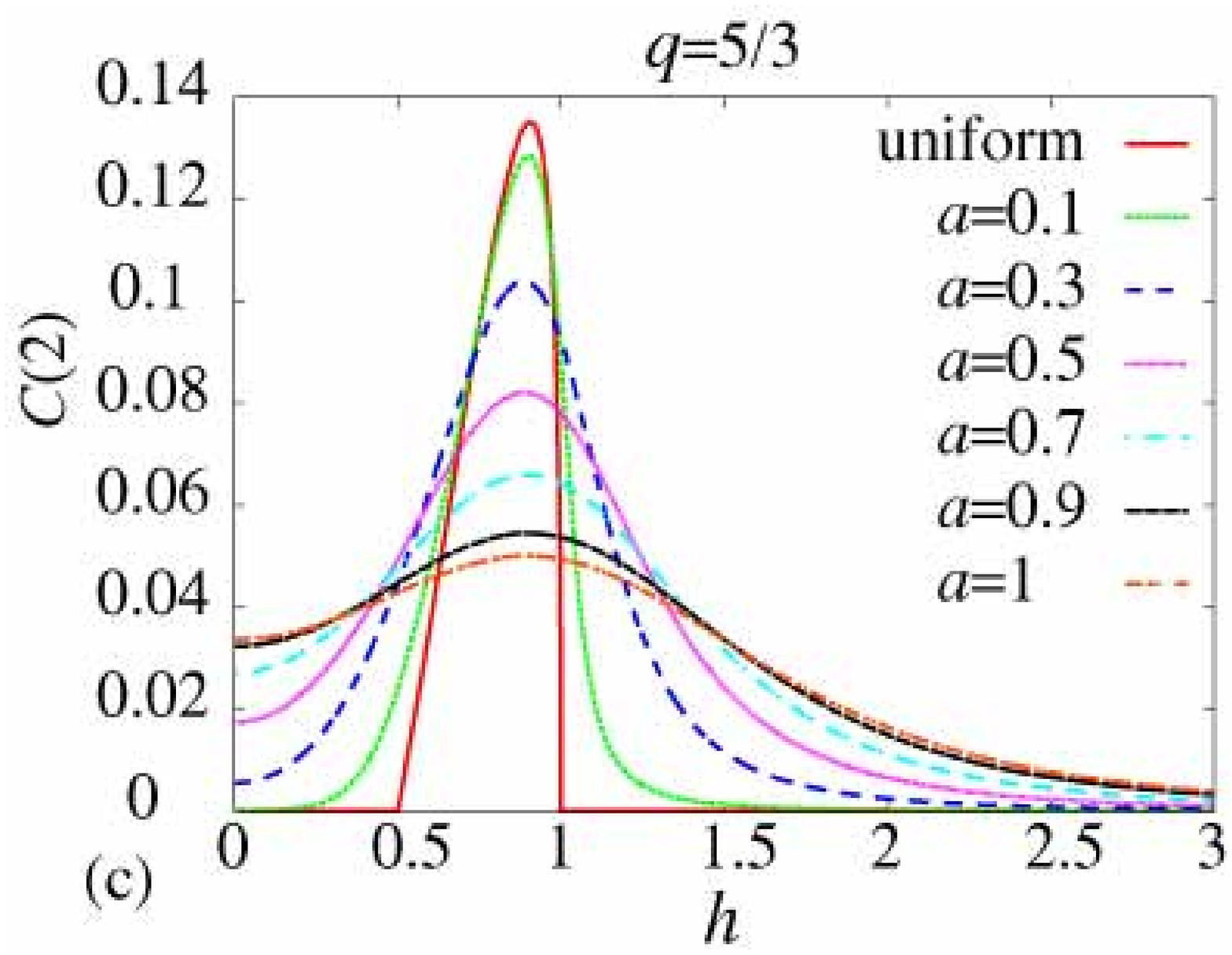}
\includegraphics[width=0.45\textwidth,clip]{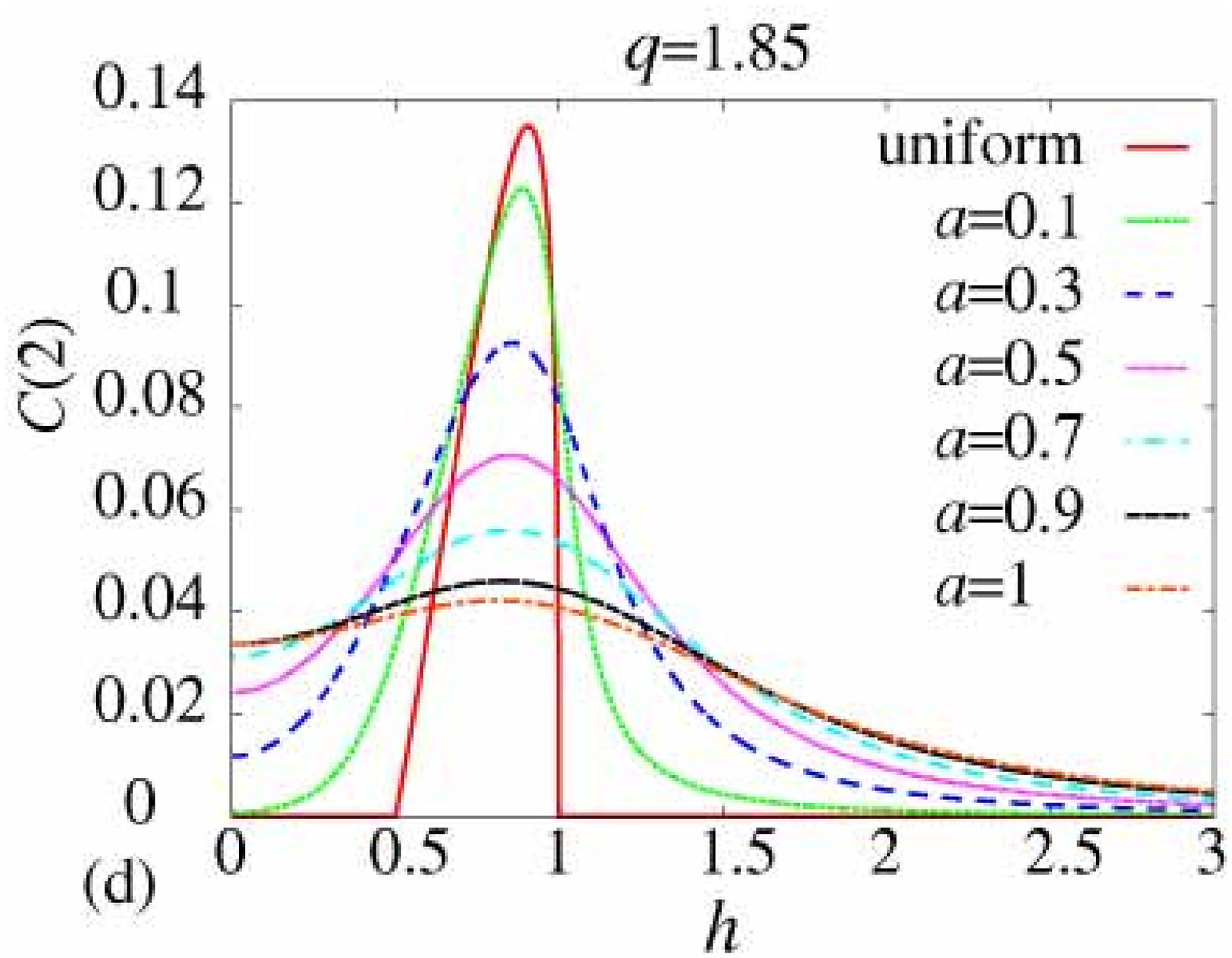}
\\
\includegraphics[width=0.45\textwidth,clip]{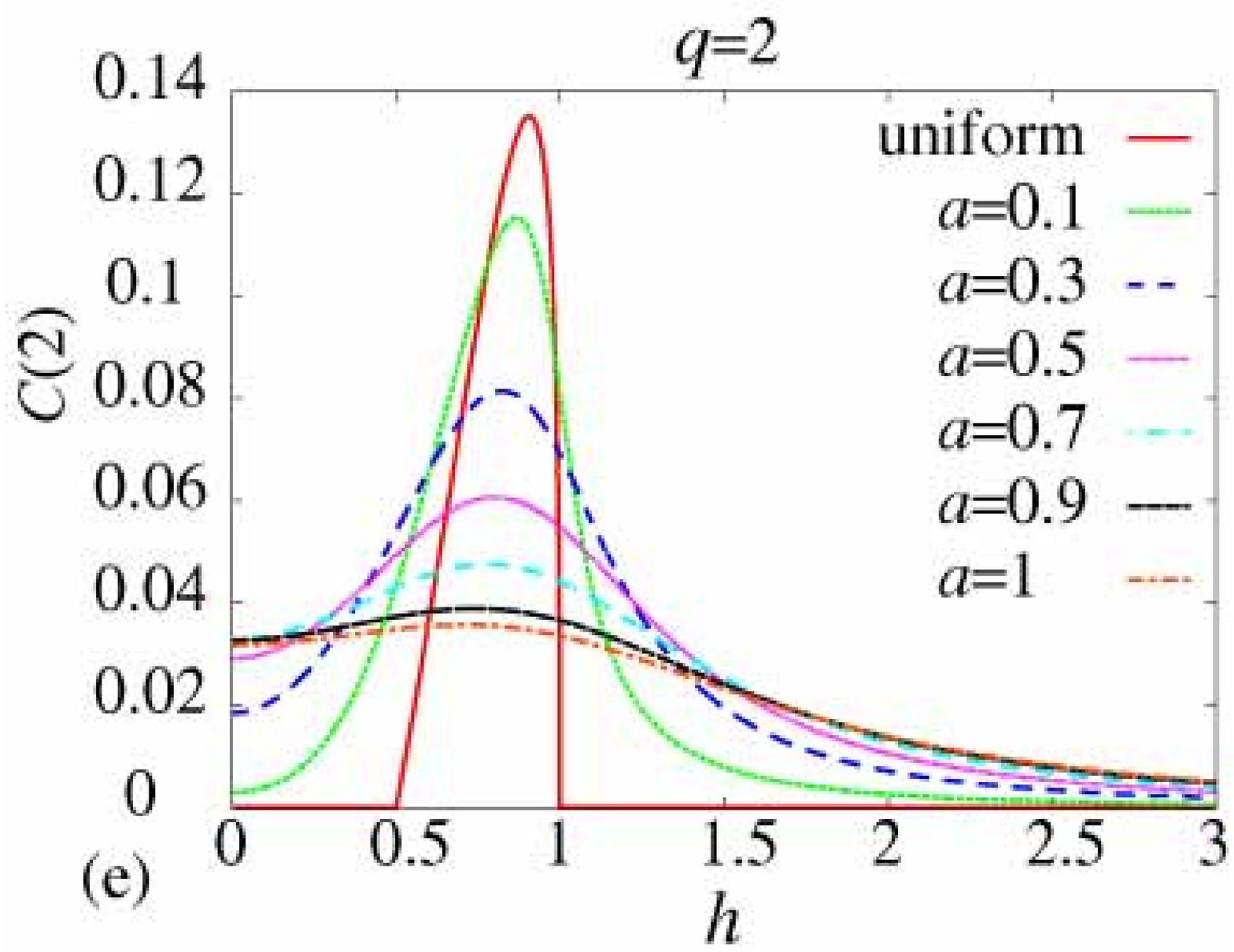}
\caption{(Color online) The next-nearest-neighbor concurrence $C(2)$ in a random magnetic field at zero temperature; (a) for $q=1$ (the Gaussian distribution); 
(b) for $q=1.35$;
(c) for $q=5/3$;
(d) for $q=1.85$;
(e) for $q=2$ (the Lorentzian distribution).
All the data are plotted as functions of the uniform magnetic field.}
\label{c2}
\end{center}
\end{figure*}
%%%%%%%%%%%%%%%%%%%%%%%%%%%%%%%%%%%%%%%%%%%%%%

As shown in Fig.~\ref{c2}, the next-nearest neighbor concurrence $C(2)$ for each $q$ is decreased for $0.5<h<1$ as the scale parameter $a$ is increased.
On the other hand, it is increased for $h<1/2$ and $h>1$ as the randomness $a$ is increased. This is in contrast to the finite-temperature case in the next subsection, where we show that the next-nearest-neighbor concurrence $C(2)$ for $h<1/2$ is \textit{not} increased at finite temperatures.

\begin{figure}
\begin{center}
\includegraphics[width=0.45\textwidth,clip]{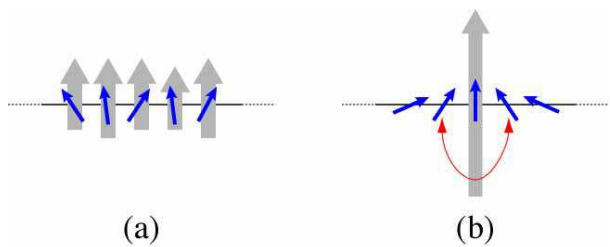}
\caption{(Color online) Smaller arrows (blue) indicate spins. Bigger arrows (grey) indicate the random magnetic field.
(a) The random magnetic field happens to be almost constant over a spatial region.
(b) The random magnetic field is extremely strong at one particular site.}
\label{image}
\end{center}
\end{figure}

The reason of the increase for $h>1$ may be the same as the above-mentioned reason for $C(1)$.
We have not been able to determine decidedly the real reason why the next-nearest-neighbor concurrence $C(2)$ is increased for $h<1/2$.
We, however, can consider some situations where the next-nearest-neighbor concurrence is increased for $h<1/2$. 
Let us first consider the case where the variance of the distribution function is finite, \textit{i.e.}~$q<5/3$.
There may arise a situation where the random magnetic field is almost constant and equal to the distribution width $a$ over a region of considerable length; see Fig.~\ref{image}(a).
%\twocolumn
%\clearpage
%%%%%%%%%%%%%%%%%%%%%%%%%%%%%%%%%%%%%%%%%%%%%%
The uniformness then may generate the concurrence in the region.
We can give an argument for this speculation.
In Fig.~\ref{c2}(a) and (b), the concurrence without the uniform field, $h=0$, is generated only for $a\geq 0.5$.
This is consistent with the fact that the concurrence without the randomness is zero for $h<0.5$;
if our speculation is correct, the concurrence is generated by an ``almost uniform'' random field only when the field is greater than 0.5.

In the case where the variance of the distribution function is infinite, \textit{i.e.}~$q\geq 5/3$, we could think of a more plausible situation.
In this case, a singularly strong random field can appear at a site as illustrated in Fig.~\ref{image}(b).
We then can take the Zeeman energy of the site as the non-perturbation term and calculate the second-order perturbation of the exchange energy. We may end up with an effective interaction between the two spins beside the strong magnetic field. In this situation, the next-nearest-neighbor concurrence may be restored around the strong field.
 
We find that the qualitative behavior of the next-nearest neighbor concurrence $C(2)$ is different depending on whether
the variance of the distribution function is finite or not.
The maximum point of the next-nearest-neighbor concurrence for $q<5/3$, where the variance of the distribution function is finite, shifts to the right as the randomness $a$ is increased as shown in Fig.~\ref{c2}(a) and (b).
In contrast, the maximum point of the next-nearest-neighbor concurrence
for $q\geq5/3$ in Fig.~\ref{c2}(c)--(e) shifts to the left as the randomness $a$ is increased. 

\begin{figure}
\begin{center}
\includegraphics[width=0.45\textwidth,clip]{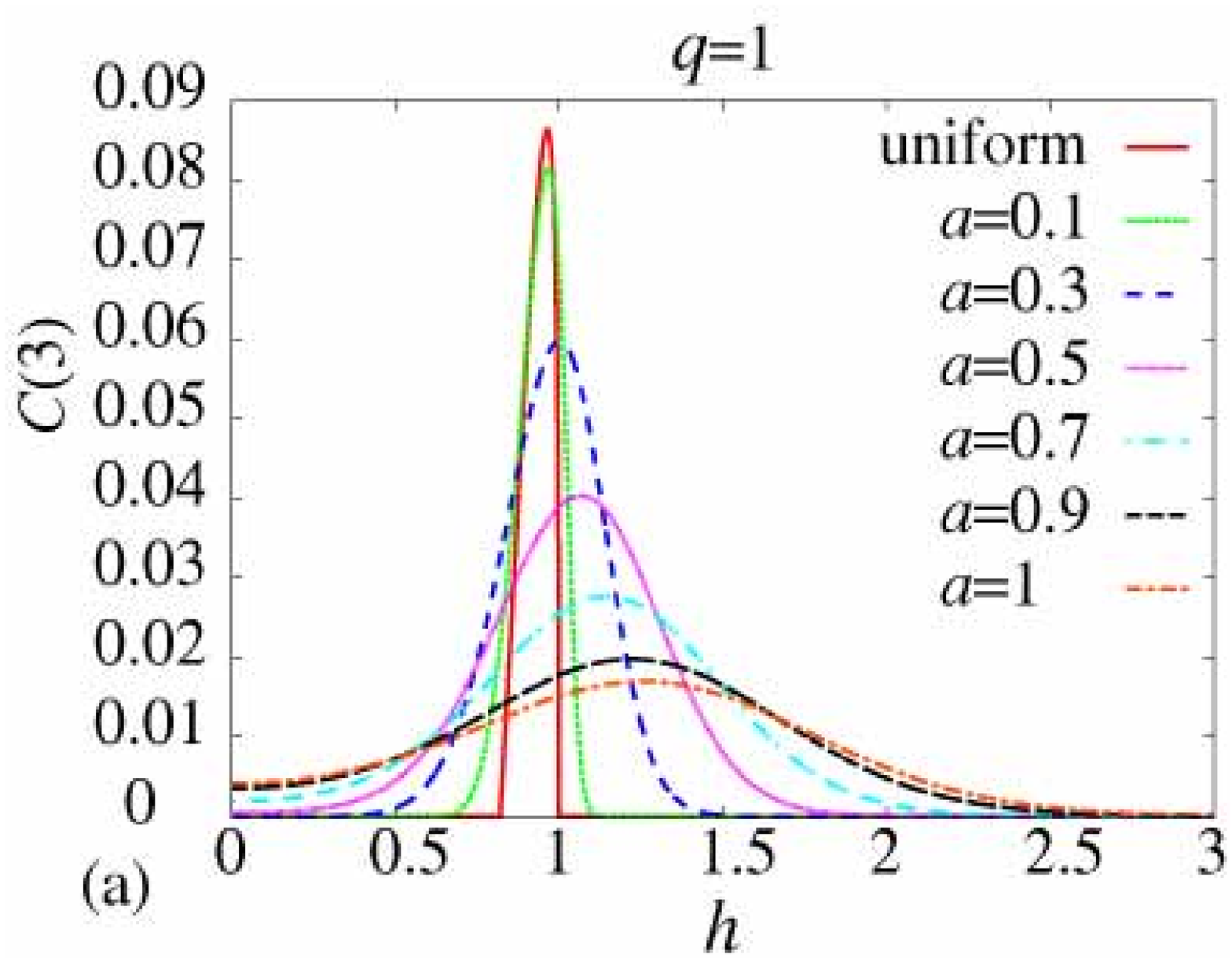} \\ 
\includegraphics[width=0.45\textwidth,clip]{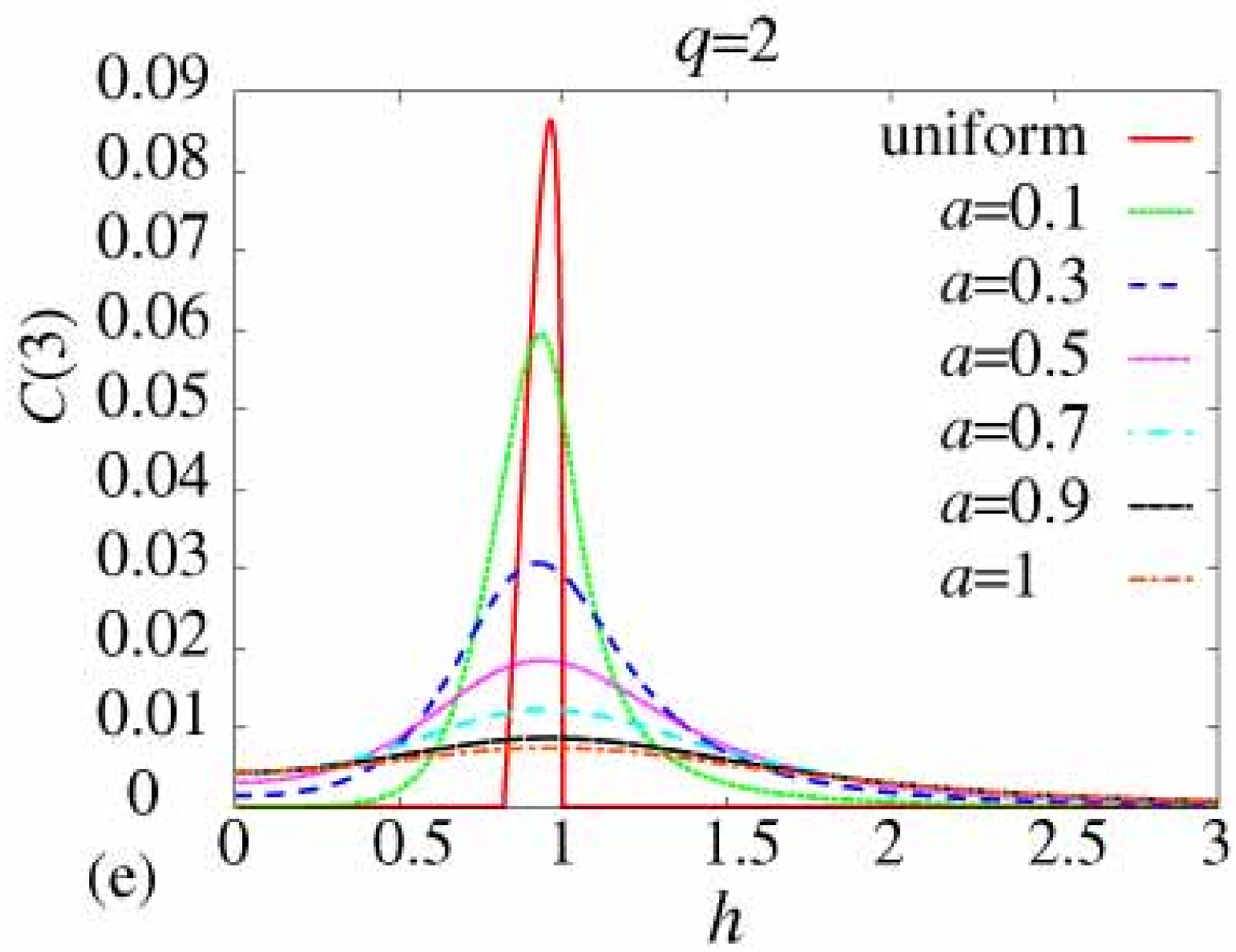}
\caption{(Color online) The third neighbor concurrence $C(3)$ in a random magnetic field at zero temperature; (a) for $q=1$ (the Gaussian distribution); 
%(b) for $q=1.35$;
%(c) for $q=5/3$;
%(d) for $q=1.85$;
(e) for $q=2$ (the Lorentzian distribution).
All the data are plotted as functions of the uniform magnetic field.}
\label{c3}
\end{center}
\end{figure}

The third-neighbor concurrence $C(3)$ and the rest, $C(4)$ and $C(5)$, behave similarly to the next-nearest-neighbor concurrence $C(2)$, only smaller than the next-nearest-neighbor concurrence. The maximum point of the third-neighbor concurrence and the rest for the cases $q\leq5/3$ first shift to the left and turn to the right as shown in Fig.~\ref{c3}.
%%%%%%%%%%%%%%%%%%%%%%%%%%%%%%%%%%%%%%%%%%%%%%
%\clearpage

\subsection{$XY$ spin chain in a uniform magnetic field at finite temperatures} \label{uni-finite-temp}
\begin{figure}
\begin{center}
\includegraphics[width=0.45\textwidth,clip]{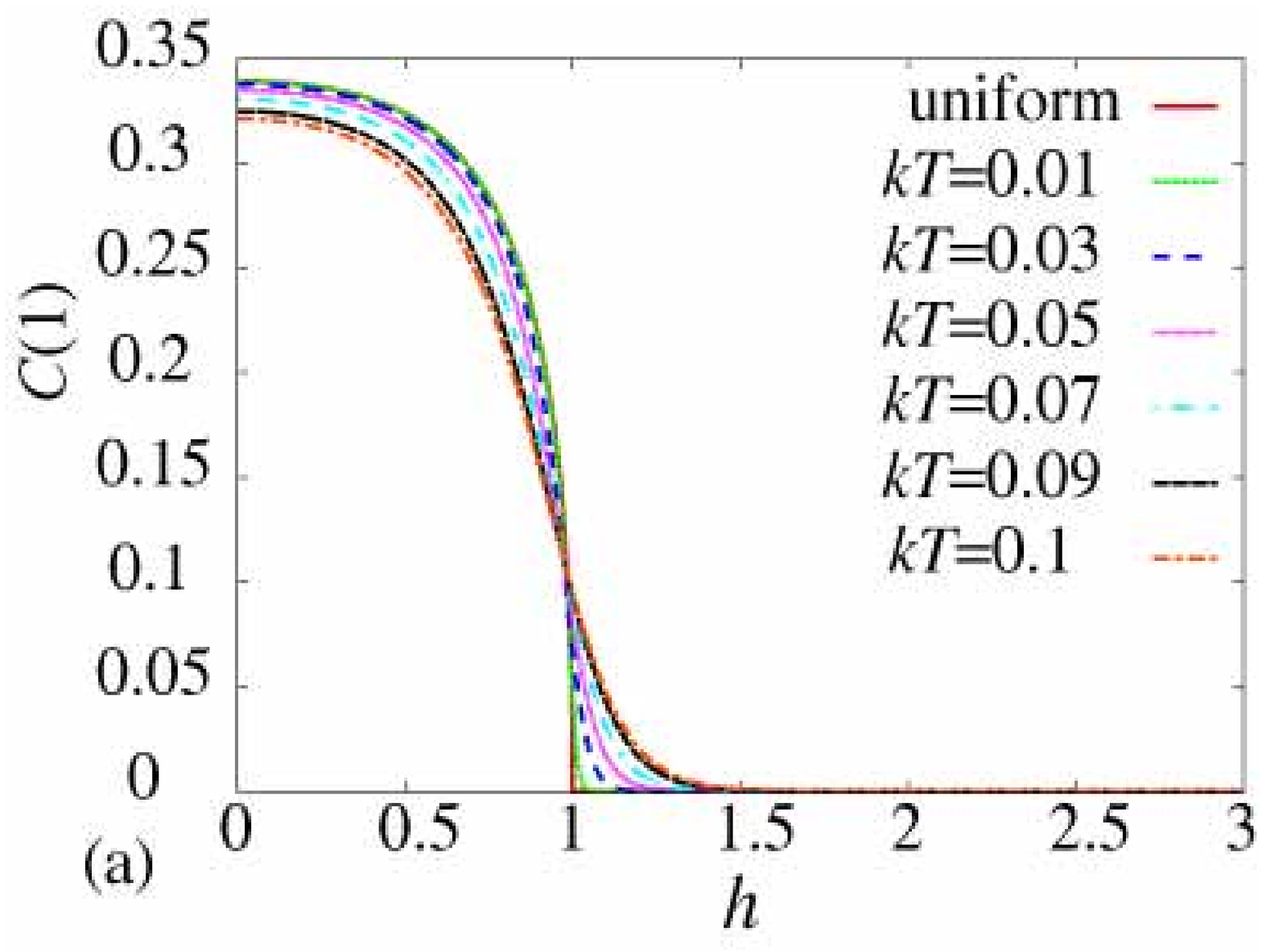} \\ 
\includegraphics[width=0.45\textwidth,clip]{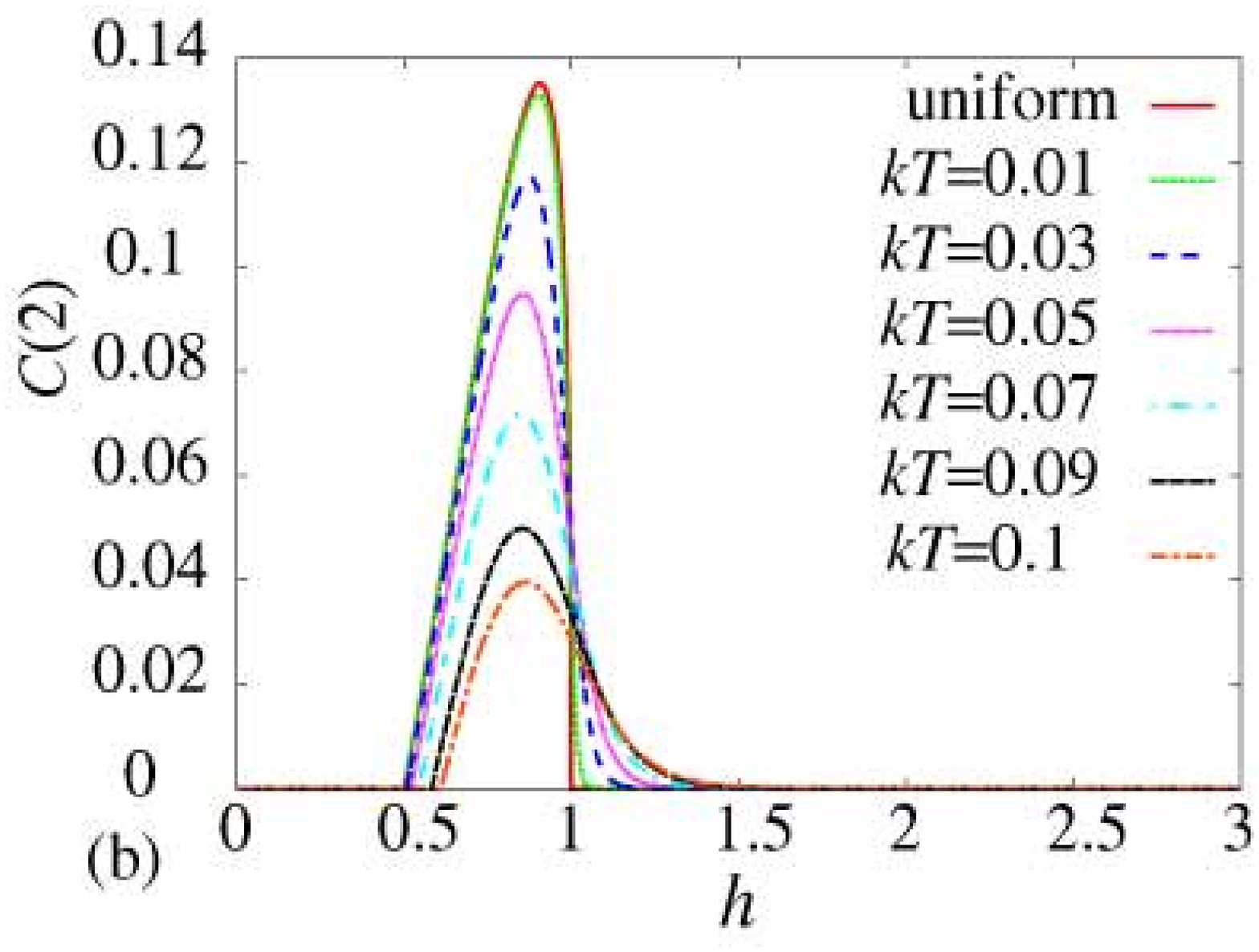}
\includegraphics[width=0.45\textwidth,clip]{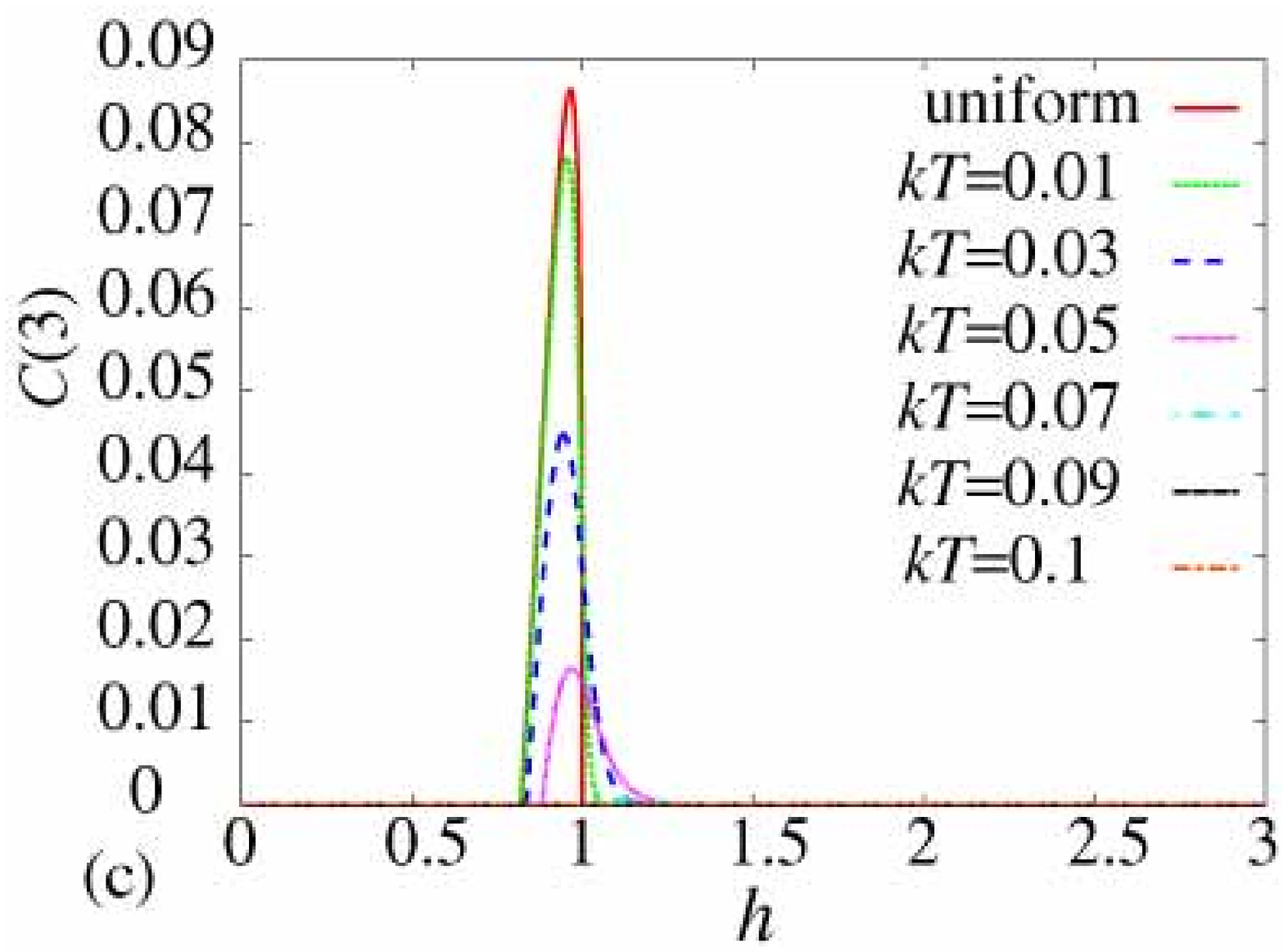}
\caption{(Color online) The concurrence in a uniform field at finite temperatures; (a) The nearest-neighbor concurrence $C(1)$; (b) The next-nearest-neighbor concurrence $C(2)$; (c) The third neighbor concurrence $C(3)$.}
\label{picthermal}
\end{center}
\end{figure}

Third, we investigate the concurrence of the $XY$ spin chain in the uniform magnetic field at finite temperatures; see Fig.~\ref{picthermal}.
%%%%%%%%%%%%%%%%%%%%%%%%%%%%%%%%%%%%%%%%%%%%%
%%%%%%%%%%%%%%%%%%%%%%%%%%%%%%%%%%%%%%
The nearest-neighbor concurrence $C(1)$ is decreased for $h<1$, whereas it is increased for $h>1$; see Fig.~\ref{picthermal}(a). The reason why the concurrence is increased for $h>1$ may be that the temperature excites entangled states above the ferromagnetic ground state with some probability due to thermal fluctuation. Thus, the concurrence can have a non-zero value for $h>1$. We can hardly see the essential difference between the effects of the random magnetic field and the temperature on the nearest-neighbor concurrence $C(1)$; compare Fig.~\ref{c1} and Fig.~\ref{picthermal}(a).

The next-nearest neighbor concurrence $C(2)$ is decreased for $0.5<h<1$. The increase of the concurrence appears only for $h>1$; the concurrence for $h<0.5$ does not appear. This is in contrast to the case of the random magnetic field shown in the previous subsection, where the next-nearest-neighbor concurrence $C(2)$ for $h<1/2$ is increased by the random field; compare Fig.~\ref{c2} and Fig.~\ref{picthermal}(b). The difference between the disturbance of the random magnetic field and the temperature appears in this point. 
The third-neighbor concurrence $C(3)$ and the rest, $C(4)$ and $C(5)$, behave similarly to the next-nearest-neighbor concurrence $C(2)$ except for quantitative difference.

\begin{figure}
\begin{center}
\includegraphics[width=0.45\textwidth,clip]{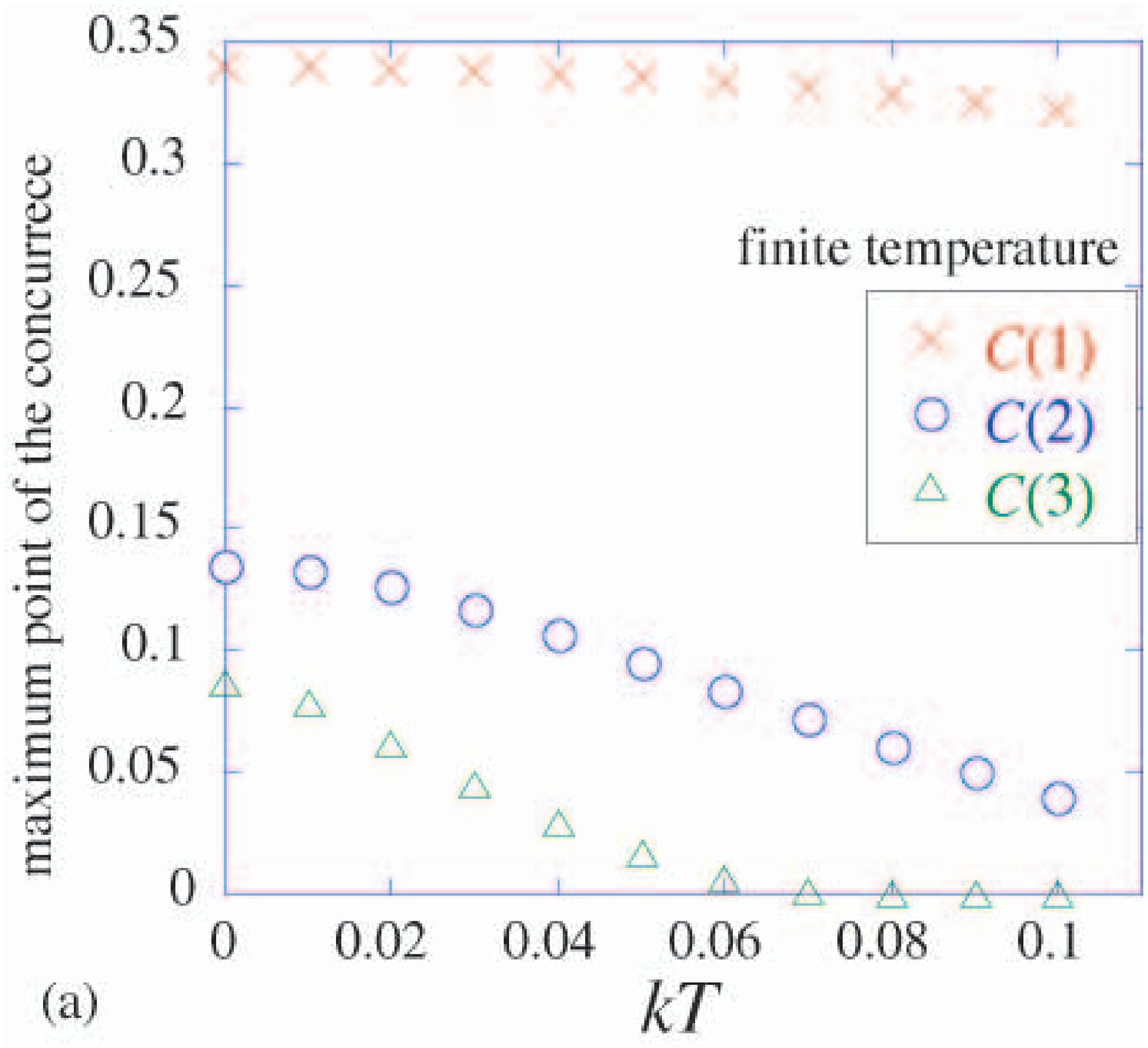}
\includegraphics[width=0.45\textwidth,clip]{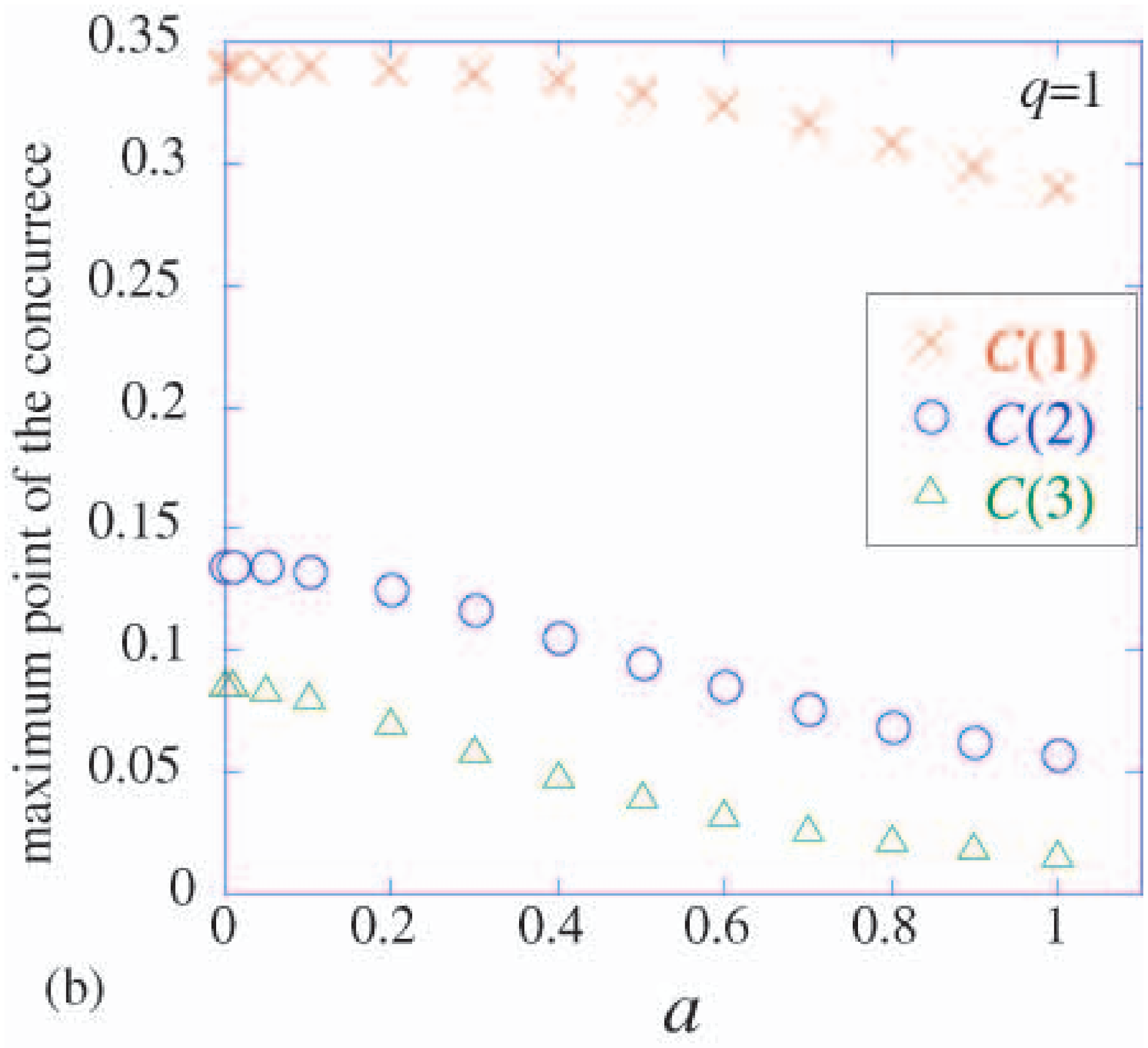}
\includegraphics[width=0.45\textwidth,clip]{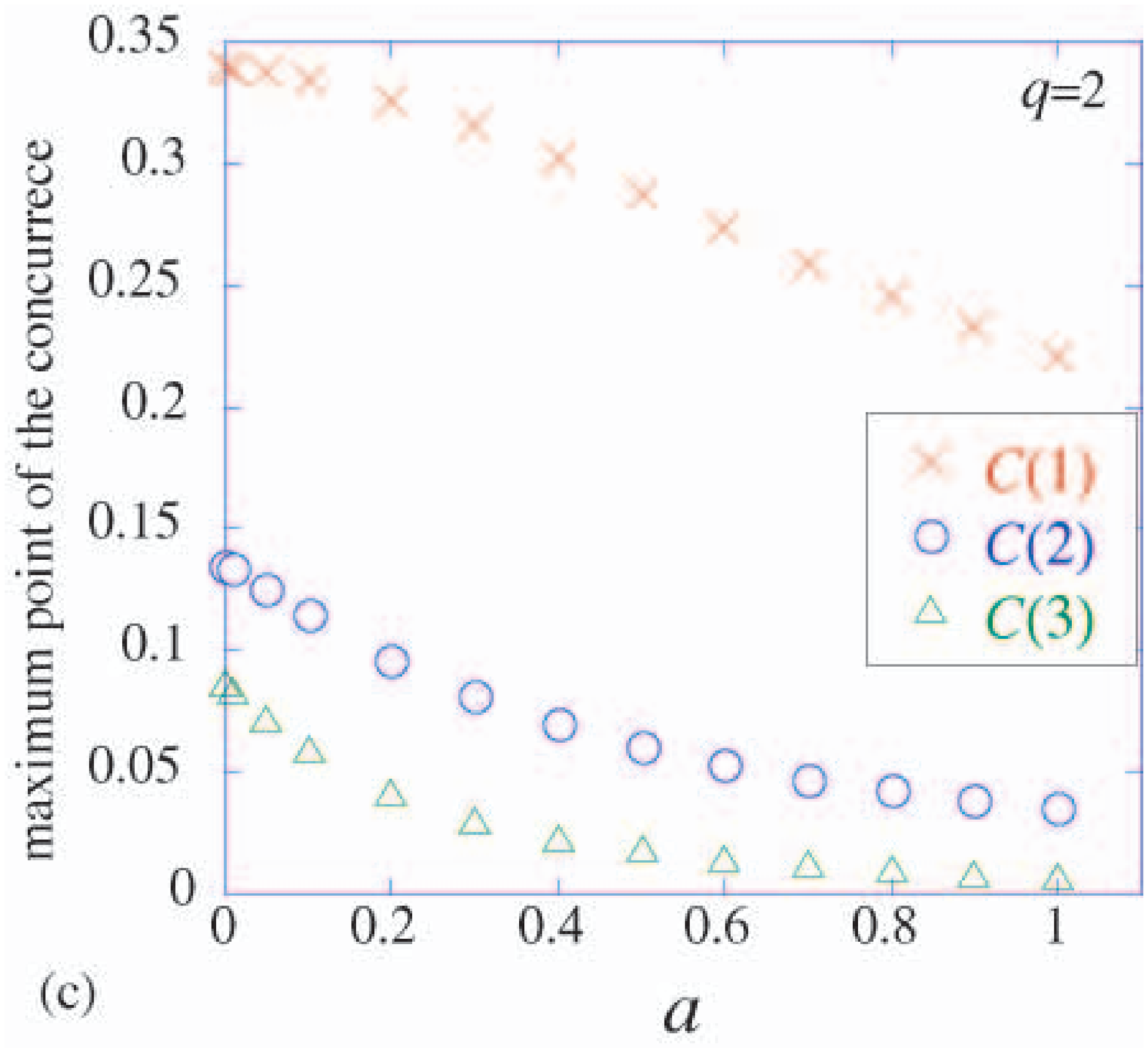}
\caption{(Color online) (a) The maximum point of the concurrence in a uniform magnetic field at finite temperatures as a function of the temperature $kT$; (b) The maximum point of the concurrence in a random magnetic field at zero temperature for $q=1$ as a function of the scale parameter $a$; (c) The maximum point of the concurrence in a random magnetic field at zero temperature for $q=2$ as a function of the scale parameter $a$.}
\label{maximum}
\end{center}
\end{figure}

Finally, we study the maximum concurrence as a function of the scale parameter $a$ or the temperature $kT$.
The reduction of the maximum concurrence as a function of the scale parameter or the temperature is plotted in Fig.~\ref{maximum}.
%%%%%%%%%%%%%%%%%%%%%%%%%%%%%%%%%%%%%%%%%%%%%
The nearest-neighbor concurrence $C(1)$ decreases for the random magnetic field more rapidly than that at finite temperatures in the plotted ranges. In the same plot ranges, however, the third-neighbor concurrence $C(3)$ in the random magnetic field remains finite, whereas the third-neighbor concurrence $C(3)$ at finite temperatures almost vanishes for $kT\geq0.07$. We thus observe that, as the distance between the two spins increases, the concurrence becomes considerably weak against the thermal fluctuation.

%%%%%%%%%%%%%%%%%%%%%%%%%%%%%%%%%%%%%%%%%%%%%%%%%%
\section{Summary and discussions} \label{part4}
We have studied the entanglement of the $XY$ spin chain in a random magnetic field at zero temperature and in a uniform field at finite temperatures.
We found that:
(i) In general, the entanglement is decreased by the random magnetic field and the temperature;
(ii) The entanglement is increased by the random magnetic field and the temperature in some parameter regions. That is, quantum correlation can be both increased and decreased by the disturbances. In particular, we
find that the next-nearest-neighbor concurrence $C(2)$ for $h<1/2$ is increased by the random magnetic field as well as for $h>1$. The increase of the concurrence for $h<1/2$ does not occur for the thermal fluctuation.
This is a notable difference between the random magnetic field and the thermal fluctuation.
%We investigate why the random magnetic field restores the entanglement.
%But we cannot find why thermal fluctuation does not restores the next-nearest-%neighbor concurrence in the region $h<1/2$.
(iii) The qualitative behavior of the maximum point of the concurrence depends on whether
the variance of the distribution function is finite or not.
In particular, the maximum point of the concurrence shifts to the right when the variance of the distribution function is finite, whereas it shifts to the left when
the variance of the distribution function is infinite.
(iv) The entanglement between two spins at finite temperatures is weaker than that in the random magnetic field at zero temperature as the length of two spin is larger.

In view of implementation of quantum information processing, the conclusion (ii) indicates that impurity and thermal disturbances are not always destructive to the entanglement resources. The conclusion (iii) also indicates that, under some randomness, we may obtain the best quality of the entanglement at a point different from the pure case.

We need further studies to find definite reasons why the concurrence is increased by the random magnetic field and the thermal fluctuation.
It may be also interesting to compute perturbationally the dependence of the concurrence on the distribution width of the random magnetic field to see why the behavior of the maximum concurrence is different for finite variance and infinite variance.

%\twocolumn
\section*{Acknowledgment}
We are grateful to Dr.~Akinori Nishino for useful suggestions and advice. 
We acknowledge support by Grant-in-Aid for Scientific Research (No.~17340115) from the Ministry of Education, Culture, Sports, Science and Technology as well as support by Core Research for Evolutional Science and Technology (CREST) of Japan Science and Technology Agency.
The use of facilities at the Supercomputer Center, Institute for Solid State Physics, University of Tokyo is gratefully acknowledged.
%\appendix
%\section{Sample}

%Equations in the appendix will be numbered as (A$\cdot$1), (A$\cdot$2), (A$\cdot$3) \ldots.

\end{document}